\documentclass[11pt,a4paper]{article}
\usepackage{amsmath,amssymb}
\usepackage{epsfig,graphicx}

\begin{document}

\title{\bf Resonant $\gamma \to a$ transition 
in magnetar magnitosphere}
\author{N.~V.~Mikheev$^a$\footnote{{\bf e-mail}: mikheev@uniyar.ac.ru},
D.~A.~Rumyantsev$^{a}$\footnote{{\bf e-mail}: rda@uniyar.ac.ru}, 
Yu.~E.~Shkol'nikova$^{a}$
\\
$^a$ \small{\em Yaroslavl State University} \\
\small{\em Yaroslavl, 150000 Russia}
}
\date{}
\maketitle

\begin{abstract}
The effect of a magnetized plasma on the resonant photoproduction of axions on the 
electromagnetic multipole components of the medium, $i \to f+a$, has been considered. 
It has been shown that the axion resonant emissivity, due to various 
reactions involving particles of the medium, is naturally
expressed in terms of the emissivity of the photon $\to$ axion transition. The number of axions 
produced by
the equilibrium cosmic microwave background radiation in the magnetar magnetosphere has been 
calculated. It has been shown that the resonant mechanism under consideration is inefficient 
for the production of cold dark mass.
\end{abstract}

\section{Introduction}

\indent\indent

The axion proposed by Peccei and Quinn~\cite{Quinn:1977} to
solve the problem of the conservation of the CP 
invariance of strong interactions remains not only the most
attractive solution to the CP problem, but also is the
most probable candidate for cold dark matter in the
Universe. Since the Peccei-Quinn symmetry 
violation scale $f_a$ is large, the interaction of axions with
matter is very weak (the coupling constant is 
$f_a^{-1} \lesssim 10^{-8}$\, GeV$^{-1}$~\cite{Raffelt:1996}). 
In view of this circumstance, the
experimental detection of the axion is a complicated
problem.

At the same time, the effect of an active 
medium on reactions involving axions can both catalyze these reactions 
and additionally (as compared to $f_a^{-1}$) suppress them, depending on the parameters of
the medium (temperature $T$, chemical potential $\mu$,
and magnetic field $B$). It is of particular interest to
analyze the processes involving axions in an extremely
strong magnetic field $B\gg B_e$
($B_e = m^2/e \simeq 4.41 \times 10^{13}$ G is the 
critical magnetic field).~\footnote{We use natural units
$c = \hbar = k = 1$, $m$ is the electron mass,  $e > 0$ is the elementary
charge.} 
 Such conditions can exist in the magnetospheres of magnetars, 
a specific class of neutron stars
with magnetic fields that are much stronger than $B_e$
and reach $10^{14} - 10^{16}$ G~\cite{Duncan:1992,Duncan:1995,Duncan:1996}. 
Moreover, a multicomponent 
(electron-positron or ion) plasma can exist
near such objects. 
In particular, electron number density in the region of closed field lines is
estimated as~\cite{Lyutikov:2002}
\begin{eqnarray}
\label{eq:n1}
n\sim \left (\frac{1}{R_{NS} \Omega} \right ) n_{GJ} \simeq 
  5\cdot 10^5 \, 
\left (\frac{10\,\mbox{rad/s}}{\Omega} \right ) 
\left (\frac{10\,\mbox{km}}{R_{NS}} \right ) n_{GJ} 
\gg n_{GJ},
\end{eqnarray}
\noindent where 
\begin{eqnarray}
\label{eq:ngj}
n_{GJ} = 7\cdot 10^{-2} \frac{B}{P}\, \mbox{cm}^{-3} \simeq 
 3\cdot 10^{13}\, \mbox{cm}^{-3} 
\left (\frac{B}{100B_e} \right )\left (\frac{10\,\mbox{s}}{P} \right ) 
\end{eqnarray}                                         
\noindent is the Goldreich-Julian charge number density~\cite{GJ:1969}.

Nowadays, there exists a growing interest to investigation of 
effective axion production processes 
in astrophysical object with such extremal conditions.  
In recently paper~\cite{Popov:2009} was considered the conversion of axions to photons in 
strong magnetic fields of neutron stars (Primakoff effect). It was  shown, that the 
photon $\to$ axion resonant conversion is absented for density plasma in 
magnetar environment 
and allowable 
axion mass $10^{-6}\lesssim m_a \lesssim 10^{-2}$ eV.
But then, the scattering processes of photons on the plasma components  
in the magnetar magnetosphere  with axion emission  become important.

\begin{figure}[h]
\centerline{\includegraphics{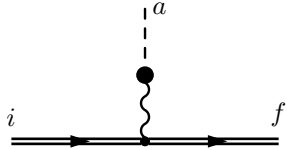}}
\caption{Feynman diagram for the general process $i \to f + a$. 
Double lines mean that the effect of the external field on
the initial and final states is exactly taken into account.}
\label{fig:itofa}
\end{figure}

\begin{figure}[h]
\centerline{\includegraphics{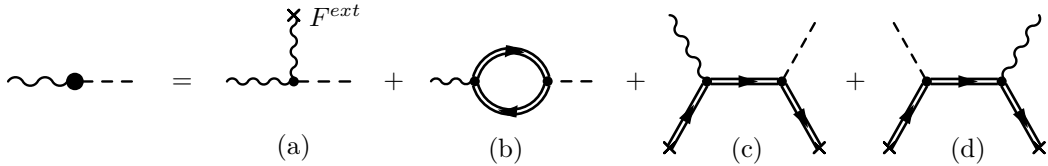}}
\caption{Feynman diagrams for the effective $\gamma a$ interaction
vertex.}
\label{fig:gagamma}
\end{figure}

Therefore, it is of interest to consider
the production of axions in the general reaction $i \to f + a$ 
(see the diagram in Fig.~\ref{fig:itofa}) under conditions very strong magnetic field 
and relatively dense plasma ($n \gg n_{GJ}$), where the initial $(i)$
and final $(f)$ states can involve the electromagnetic
multipole components of the medium. The closed circle in Fig.~\ref{fig:itofa} is the 
effective $g_{\gamma a}$ interaction vertex (diagrams in Fig.~\ref{fig:gagamma}). 
It is easy to see that the process under
consideration can be resonant owing to the presence
of a virtual photon. Such process was considered recently in~\cite{MRS:2009} and 
our investigation bases on results of this paper. 
A similar situation for the region
close to resonance was recently considered also in~\cite{Skobelev:2007} for
the Compton scattering of relic photons on the electrons and positrons 
of the magnetar magnetosphere.
However, we will show below that the results obtained
in~\cite{Skobelev:2007} are inaccurate.

\section{Amplitude of process $i \to f + a$}

In the existing axion models the process $i \to f + a$ in the presence of 
the external magnetic field 
can be described by the effective Lagrangian~\cite{Raffelt:1996}
\begin{eqnarray}
\label{eq:L1}
{\cal L}_{a\gamma}(x) &=& g_{a\gamma}\widetilde F^{\mu \nu} [\partial_{\nu}A_{\mu}(x)] a(x) + 
\frac{g_{af}}{2m_f} 
[\bar \psi_f(x) \gamma^{\mu} \gamma_5 \psi_f(x)] \partial_{\mu} a(x) - 
\\
\nonumber
&-&Q_f[\bar \psi_f(x) \gamma^{\mu} \psi_f(x)]A_{\mu}(x).
\end{eqnarray}
\noindent Here $A_{\mu}$ is the four-potential of the quantized electromagnetic field, 
$\widetilde F^{\mu \nu}$ is the dual tensor of the external field,  $\psi_f(x)$  and $a(x)$  
are the quantized fermion and axion fields, respectively, 
  $g_{a\gamma} = \alpha \xi/2\pi f_a$, where $\xi$ is the
model dependent parameter about unit,  
$g_{af} = C_f m_f/f_a$ is the dimensionless Yukawa axion-fermion coupling
constant with the model dependent factor $C_f$, $Q_f$ is
the electric charge of the fermion (for the electron $Q_f = - e$).

Using Lagrangian~(\ref{eq:L1}), the $i \to f + a$ amplitude can
be represented in the form
\begin{equation}
{\cal M}^a_{i \to f} 
 = - \frac{{\cal M}^{\gamma}_{if} {\cal M}_{\gamma \to a}}
{q'^2 -{\cal P}^{(\varepsilon)}(q')}\, , 
\label{eq:M1}                                                       
\end{equation}
\noindent where ${\cal M}^{\gamma}_{if}$ is the amplitude of the process $i \to f + \gamma$
with the emission of a photon in the final state;
\begin{equation}
{\cal M}_{\gamma \to a} 
 = i \bar g_{a\gamma} (\varepsilon \widetilde F q') 
\label{eq:M2}                                                       
\end{equation}
\noindent is the photon $\to$ axion transition amplitude; 
$q'^{\mu} = (\omega',{\bf k}')$ is the axion four-momentum; and 
${\cal P}^{(\varepsilon)}(q')$ is the
eigenvalue of the photon polarization operator, which
corresponds to the polarization vector $\varepsilon_{\alpha}$. The effective
axion-photon coupling constant $\bar g_{a\gamma}$ 
can be represented as the sum of three terms $\bar g_{a\gamma} = g_{a\gamma} + 
\Delta g^{B}_{a\gamma} + \Delta g^{pl}_{a\gamma}$. 
The first term corresponds to the interaction of
the axion with the electromagnetic field caused by the
Adler anomaly (the diagram in Fig.~\ref{fig:gagamma}a), the second
term describes the axion-photon interaction through
an electron loop (the diagram in Fig.~\ref{fig:gagamma}b), and the third
term corresponds to forward scattering on the electrons and positrons of the plasma 
(the diagrams in Figs.~\ref{fig:gagamma}c and~\ref{fig:gagamma}d). 
A similar calculation of $\Delta g^{B}_{a\gamma}$ 
and $\Delta g^{pl}_{a\gamma}$ 
was previously performed in~\cite{Mikheev:1999} and~\cite{Mikheev:2006}, respectively.
Here, we only note that to correctly calculate $\Delta g^{B}_{a\gamma}$, the
subtraction corresponding to the Adler anomaly
should be made in it~\cite{Mikheev:1999}. In particular, this fact was 
disregarded in~\cite{Skobelev:2007}, which is one of the causes of why the
results obtained in that work are incorrect.

Further, we represent ${\cal P}^{(\varepsilon)}(q')$ in the form 
${\cal P}^{(\varepsilon)} = \Re - i\Im$, where $\Re = Re ({\cal P})$ 
and $\Im =  Im ({\cal P})$ are the
real and imaginary parts of the polarization operator,
respectively. The latter is due to the processes of the
absorption and emission of photons in the plasma and,
according to~\cite{Weldon:1983}, is expressed as
\begin{eqnarray}
\label{eq:I1}                                                       
\Im = \omega' \left (e^{\omega'/T} - 1 \right ) \Gamma_{cr} ,
\end{eqnarray}
\noindent in terms of the total photon production width
\begin{eqnarray}
\label{eq:II1}                                                       
\Gamma_{cr} = \sum_{i,f} \int  |{\cal M}^{\gamma}_{if}|^2 d\Phi_{if} ,
\end{eqnarray}
\noindent  where $d\Phi_{if}$ is the phase volume element of the states
$i$ and $f$ for the process $i \to f+\gamma$ taking into account the corresponding 
distribution functions, and summation is
performed over all of the possible initial and final
states.

\section{Axion emissivity}

Taking into account the above consideration, the
axion emissivity (the loss of energy
from a unit volume per unit time due to the
axion escape), due to the reactions involving the
particles of the plasma, can be represented in the form
\begin{equation}
Q = \sum_{i,f} \int d\Phi_{if} \, d\Phi' \, \omega' 
|{\cal M}^{a}_{i \to f}|^2 \, ,
\label{eq:Qa0}                                                       
\end{equation}
\noindent $d\Phi' = \frac{d^3 k'}{(2\pi)^3 2 \omega'}$  is the phase volume element of the axion.
Taking into account Eqs.~(\ref{eq:M1}) and~(\ref{eq:I1}), 
$Q$ is represented in the form
\begin{equation}
Q = \int  \frac{d\Phi' \, |{\cal M}_{\gamma \to a}|^2}{e^{\omega'/T} - 1} \, 
\frac{\Im
} 
{(q'^2 -\Re)^2 + \Im^2} \, .
\label{eq:Qa1}                                                       
\end{equation}
\begin{figure}[h]
\centerline{\includegraphics{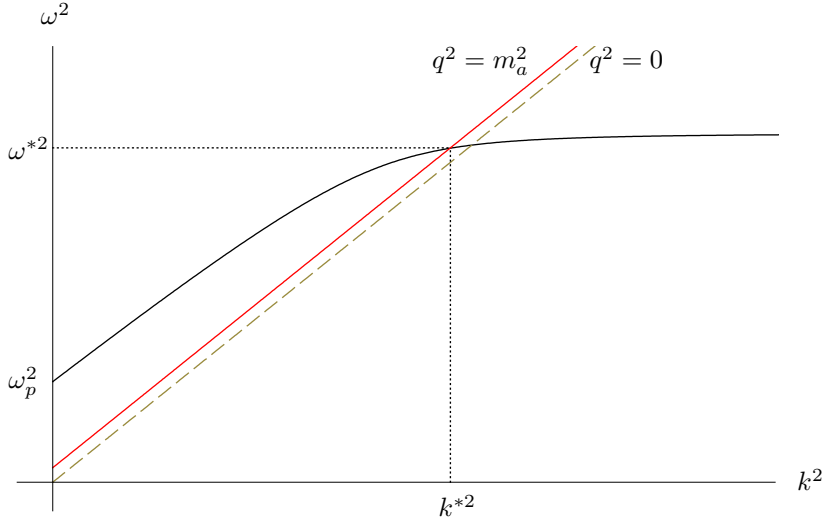}}
\caption{Photon (black curve) and axion (red line) dispersion law. 
The crosspoint of the photon and axion dispersion 
curves corresponds to resonant point.}
\label{fig:disp}
\end{figure}

According to Eq.~(\ref{eq:Qa1}), the most significant 
contribution to the axion luminosity comes from the 
resonance region, i.e., from the vicinity of the intersection
point of the dispersion curves of the axion, $q^{\, \prime 2} = m_a^2$, 
and photon, $q^{\, \prime 2} = \Re$, so that the photon becomes real (Fig.~\ref{fig:disp}).
Near the resonance, the part of the integrand in
Eq.~(\ref{eq:Qa1}) can be interpolated by the $\delta$ function
\begin{equation}
 \frac{\Im}
{(q'^2 -\Re)^2 + \Im^2} \simeq \pi \, \delta(q'^2 -\Re) \, .
\label{eq:Del1}
\end{equation}
\noindent
Using the properties of the $\delta$ function, we can represent 
Eq.~(\ref{eq:Qa1}) in the form
\begin{equation}
 \frac{\Im}
{(q'^2 -\Re)^2 + \Im^2} \simeq \pi \, 
\int \frac{d^3 k}{2 \omega} Z_{\varepsilon} \delta^4 (q-q') \, ,
\label{eq:Del2}
\end{equation}
\noindent
 where $Z_{\varepsilon}^{-1} = 1-\frac{\partial \Re}{\partial \omega^2}$ 
corresponds to the renormalization of the photon wavefunction, 
$q^{\mu} = (\omega,{\bf k})$ is the photon four-momentum.

Taking into account Eq.~(\ref{eq:Del2}), emissivity~(\ref{eq:Qa1}) can be
written as

\begin{eqnarray}
\label{eq:Qa2}
Q \simeq (2\pi)^4 \, 
\int \frac{d^3 k}{2 \omega (2\pi)^3} \,\frac{\omega}{e^{\omega/T} - 1} 
 \int \frac{d^3 k'}{2 \omega' (2\pi)^3}\, Z_{\varepsilon} 
|{\cal M}_{\gamma \to a}|^2 \delta^4 (q-q') \, .
\end{eqnarray}

\noindent 
This expression exactly corresponds to the formula for
the axion emissivity in the $\gamma \to a$ process. Thus, the
axion emissivity in the resonance region, owing to the
reactions involving the particles of the medium, is naturally 
expressed in terms of the photon $\to$ axion
transition emissivity.

After integration with $\delta$ functions, emissivity is
reduced to the form
\begin{eqnarray}
Q =  \frac{\bar g_{a\gamma}^2 (eB)^2}{32 \pi^2 \alpha} \, 
\int_{-1}^1 \frac{dx}{e^{\omega/T}-1} \, 
\frac{Z_{\varepsilon} k 
(\varepsilon \widetilde \varphi q)^2}{\left |1-\frac{d\omega^2}{dk^2}\right |}\bigg |_{k=k^*}\, .
\label{eq:Qa3}
\end{eqnarray}

\noindent Here $x = \cos{\theta}$, where $\theta$  is the angle between 
the photon momentum and the magnetic field, $k^* = k^*(\theta)$ is a
root of the equation  $\omega^2 (\vec k) = m_a^2+k^2$,  
 $\widetilde \varphi_{\alpha \beta} = \widetilde F_{\alpha \beta}/B$.

The further calculation of the emissivity significantly 
depends on the plasma characteristics finally
determining the dispersion properties of photons. We
consider below two particular cases.

i) Weakly magnetized dense plasma, $m_a^2 \ll eB \ll T^2 , \mu^2$. 
In this case, $\varepsilon_\alpha$ is the polarization vector of the
longitudinal plasmon
\begin{eqnarray}
\varepsilon_\alpha = \sqrt{\frac{q^2}{(uq)^2-q^2}}\, 
\left (u_\alpha - \frac{(uq)}{q^2}\, q_\alpha \right ) , 
\end{eqnarray}
\noindent where $u_\alpha$ is the four velocity of the plasma. 
Emissivity~(\ref{eq:Qa3})  has the simple form
\begin{eqnarray}
Q =  \frac{\bar g_{a\gamma}^2 (eB)^2}{48 \pi^2 \alpha} \, 
 \frac{(k^*)^3}{e^{k^*/T}-1} \, 
\label{eq:Qa4}
\end{eqnarray}
\noindent in complete agreement with the results 
reported in~\cite{MRV:1998}. Note that $k^*$ in this case is independent of 
$\theta$ and
is determined only by the plasma parameters.

ii) Strongly magnetized plasma, $eB \gg m^2 ,\, \mu^2 \gg T^2$. Here 
$\varepsilon_\alpha = 
(q \widetilde \varphi)_\alpha/\sqrt{(q \widetilde \varphi \widetilde \varphi q)}$, 
$\Re \simeq (q \widetilde \varphi \widetilde \varphi q) \left (\frac{\omega_p^2(1+\eta)}{\omega^2} -
\eta \right )$, $\eta = (\alpha/3\pi)(B/B_e)$ and plasma frequency $\omega_p$ 
is related to the electron density as $\omega^2_p = 4\pi \alpha n/m$.
Moreover, in the strong magnetic field limit 
the contributions of diagrams b, c and d in Fig.~\ref{fig:gagamma} in the effective
axion-photon coupling constant are suppressed by field, 
so that $\bar g_{a\gamma} \simeq g_{a\gamma}$. 

However, in contrast to the case of weakly
magnetized plasma, the final analytical expression for
emissivity can be obtained only in some particular
cases.

\begin{itemize} 

\item When the axion mass is the smallest parameter
of the problem, i.e., $\omega_p, \,T \gg m_a$ (e.g., the production
of light axions (axion mass smaller than $10^{-5}$ eV) in the
magnetar magnetosphere (see Eq.~(\ref{eq:n1})), $k^* \simeq \omega_p \sqrt{1+1/\eta}$ 
and emissivity can be represented in the
form
\begin{eqnarray}
\label{eq:Qa5a}
Q \simeq  \frac{g_{a\gamma}^2 (eB)^2}{16 \pi^2 \alpha} \,
\omega_p^3 \frac{(1+\eta)^{3/2}}{\eta^{5/2}} \left (\exp{\left [\frac{\omega_p}{T} 
\sqrt{1+\frac{1}{\eta}}\, \right ]}-1 \right )^{-1} \, .
\end{eqnarray}

\item  When $\omega_p \gg T \sim m_a$, the integral in Eq.~(\ref{eq:Qa3}) is
accumulated in the region $x \simeq 1$ and, hence, $k^* \simeq \omega_p $. In
this case, emissivity is suppressed exponentially by plasma density
%
\begin{eqnarray}
\label{eq:Qa5b}
Q &\simeq&  \frac{g_{a\gamma}^2 (eB)^2}{16 \pi^2 \alpha} \, T m_a^2  \, 
e^{-\omega_p/T} \, .
\end{eqnarray}

\end{itemize}


In addition to emissivity, it is also of interest to
estimate the number of axions produced in the magnetar magnetosphere in 
unit volume per unit time
through the above resonant mechanism, because the
axion is one of the main candidates for the constituents of cold dark matter. 
Similar to Eqs.~(\ref{eq:Qa3}),~(\ref{eq:Qa5a}), and~(\ref{eq:Qa5b}), we obtain
\begin{eqnarray}
\frac{dN}{dt dV} =  \frac{g_{a\gamma}^2 (eB)^2}{32 \pi^2 \alpha} \, 
\int_{-1}^1 \frac{dx}{e^{\omega/T}-1} \, 
\frac{k Z_{\varepsilon} 
(\varepsilon \widetilde \varphi q)^2}{\omega 
\left |1-\frac{d\omega^2}{dk^2}\right |}\bigg |_{k=k^*}\, ,
\label{eq:Na0}
\end{eqnarray}
\begin{eqnarray}
\label{eq:Na1}
\frac{dN}{dt dV} \simeq  \frac{g_{a\gamma}^2 (eB)^2}{16 \pi^2 \alpha} \,
\omega_p^2 \frac{1+\eta}{\eta^{2}}\times
 \left (\exp{\left [\frac{\omega_p}{T} 
\sqrt{1+\frac{1}{\eta}}\, \right ]}-1 \right )^{-1} \, , \quad \omega_p,\, T \gg m_a,
\end{eqnarray}
\begin{eqnarray}
\label{eq:Na2}
\frac{dN}{dt dV} &\simeq&  \frac{g_{a\gamma}^2 (eB)^2}{16 \pi^2 \alpha} \,
\frac{T m_a^2}{\omega_p}  \, e^{-\omega_p/T} \, , \quad \omega_p \gg T \sim m_a.
\end{eqnarray}

In particular, for the number of axions produced by
the cosmic microwave background radiation ($T \sim m_a \sim 10^{-3}$ eV), 
the minimum plasma density ($\sim 10^{15}$\,cm$^{-3}$) at
which the resonant mechanism is allowed  
$(\omega_p \gtrsim  m_a)$, 
and the magnetic field $B = 100B_e$, Eq.~(\ref{eq:Na0}) gives
the estimate $dN/(dV dt)\sim 10^{10}$ axions in cm$^{-3}$ per second.
Thus, $10^{29}$ axions are produced per second in the volume of the magnetar 
magnetosphere ($\sim 10^{19}$\,cm$^3$) with
the strong magnetic field. In the most optimistic variant, estimating the 
number of magnetars in the Galaxy
as $\sim 10^6$, we conclude that they produce $\sim 10^{51}$ axions in
$\sim 10^9$ yr; therefore, the density of axions in the Galaxy
should be $n_a \sim 10^{-21}$ cm$^{-3}$, which is much lower than
the density of baryons $n_b \sim 10^{-7}$ cm$^{-3}$ $\gg n_a$. 

Consequently, the statement 
made in~\cite{Skobelev:2007} that "vicinities of
magnetic neutron stars with fields $B \gg B_e$ can be high 
power generators conversing the cosmic microwave
background radiation to the axion component of the
cold dark mass" is invalid.

\section{Conclusion}

We have considered the resonant photoproduction of axions 
in the general reaction $i \to f + a$. It has been shown that the calculation of axion
emissivity owing to this process is reduced to the calculation of the emissivity of 
the photon $\to$ axion transition. Two particular cases of weakly and strongly
magnetized plasmas have been analyzed. The number
of axions produced by the equilibrium cosmic microwave background radiation in the 
magnetar magnetosphere has been estimated. It has been shown that
this mechanism is inefficient for the production of cold
dark mass even at the plasma density $n \sim 10^{15}$ cm $^{-3}$.

\bigskip

{\bf Acknowledgements}  

We express our deep gratitude to the organizers of the 
Seminar ``Quarks-2010'' for warm hospitality.

This work was performed in the framework of realization of the Federal
Target Program
``Scientific and Pedagogic Personnel of the Innovation Russia'' for 2009 -
2013 (project no. NK-410P-69)
and was supported in part by the Ministry of Education
and Science of the Russian Federation under the Program
``Development of the Scientific Potential of the Higher
Education'' (project no. 2.1.1/510).



\end{document}